\documentclass{aa}  

\usepackage{graphicx}
\usepackage{txfonts}
\newcommand{\HI}{H\,{\sc{i}}} 
\newcommand{\MHI}{$\mathrm{M_{HI}}$}
\newcommand{\DHI}{$\mathrm{D_{HI}}$}
\newcommand{\Msun}{$\mathrm{M_{\odot}}$} 
\newcommand{\Mstellar}{$\mathrm{M_{\ast}}$}
\newcommand{\kms}{km\,s$^{-1}$}
\newcommand{\omegahi}{$\mathrm{\Omega_{HI}}$}
\newcommand{\uJy}{$\mu$Jy}

\begin{document} 

\title{MIGHTEE-HI: The H\,{\sc{i}} emission project of the MeerKAT
  MIGHTEE survey\thanks{Raw data available at https://archive.sarao.ac.za}}

\author{Natasha Maddox\inst{1}
          \and
          Bradley S. Frank\inst{2,}\inst{3,}\inst{4}
          \and
          A. A. Ponomareva\inst{5}
          \and
          M. J. Jarvis\inst{5,}\inst{6}
          \and
          E. A. K. Adams\inst{7,}\inst{8}
          \and
          R. Dav{\'e}\inst{9,}\inst{6,}\inst{10}
          \and
          T. A. Oosterloo\inst{7,}\inst{8}
          \and
          M. G. Santos\inst{6,}\inst{2}
          \and
          S. L. Blyth\inst{4}
          \and
          M. Glowacki\inst{11}
          \and
          R. C. Kraan-Korteweg\inst{4}
          \and
          W. Mulaudzi\inst{4}
          \and
          B. Namumba\inst{12}
          \and
          I. Prandoni\inst{13}
          \and
          S. H. A. Rajohnson\inst{4}
          \and
          K. Spekkens\inst{14}
          \and
          N. J. Adams\inst{5}
          \and
          R. A. A. Bowler\inst{5}
          \and
          J. D. Collier\inst{3,}\inst{15}
          \and
          I. Heywood\inst{5,}\inst{12,}\inst{2}
          \and
          S. Sekhar\inst{3,}\inst{16}
          \and
          A. R. Taylor\inst{3,}\inst{11}
          }

\institute{Faculty of Physics, Ludwig-Maximilians-Universit\"at,
Scheinerstr. 1, 81679 Munich, Germany\\
\email{natasha.maddox@gmail.com}
\and
South African Radio Astronomy Observatory, 2 Fir Street,
Observatory, 7925, South Africa
\and
The Inter-University Institute for Data Intensive Astronomy
(IDIA), and University of Cape Town, Private Bag X3, 
Rondebosch, 7701, South Africa
\and
Department of Astronomy, University of Cape Town, Private Bag
X3, Rondebosch 7701, South Africa
\and
Oxford Astrophysics, Denys Wilkinson Building,
University of Oxford, Keble Rd, Oxford, OX1 3RH, UK
\and
Department of Physics and Astronomy, University of the Western Cape,
Robert Sobukwe Road, Bellville 7535, South Africa
\and
ASTRON, the Netherlands Institute for Radio Astronomy, Oude
Hoogeveesedijk 4,7991 PD Dwingeloo, The Netherlands
\and
Kapteyn Astronomical Institute, PO Box 800, 9700 AV Groningen,
The Netherlands
\and
Institute for Astronomy, University of Edinburgh, EH9 3HJ, UK
\and
South African Astronomical Observatory, PO Box 9, Observatory, 
7935, South Africa
\and
The Inter-University Institute for Data Intensive Astronomy
(IDIA), and University of the Western Cape,
Robert Sobukwe Road, Bellville 7535, South Africa
\and
Department of Physics and Electronics, Rhodes University, PO
Box 94, Makhanda, 6140, South Africa
\and
INAF - Istituto di Radioastronomia, Via P. Gobetti 101, 40129
Bologna, Italy
\and
Department of Physics and Space Science, Royal Military College
of Canada, PO Box 17000, Station Forces, Kingston, Ontario, Canada K7K 7B4
\and
School of Science, Western Sydney University, Locked Bag 1797,
Penrith, NSW 2751, Australia
\and
National Radio Astronomy Observatory, 1003 Lopezville Road,
Socorro, NM 87801, USA
}

\date{Received October 13, 2020; accepted November 10, 2020}

\abstract{We present the \HI\ emission project within the
  MIGHTEE survey, currently being carried out with the newly
  commissioned MeerKAT radio telescope. This is one of the first deep,
  blind, medium-wide interferometric surveys for neutral hydrogen
  (\HI) ever undertaken, extending our knowledge of \HI\ emission to
  $z=0.6$. The science goals of this medium-deep, medium-wide survey
  are extensive, including the evolution of the neutral gas content of
  galaxies over the past 5 billion years. Simulations predict nearly
  3000 galaxies over $0<z<0.4$ will be detected directly in \HI, with
  statistical detections extending to $z=0.6$. The survey allows us to
  explore \HI\ as a function of galaxy environment, with massive
  groups and galaxy clusters within the survey volume. Additionally, 
  the area is large enough to contain as many as 50 local galaxies
  with \HI\ mass $<10^8$\,\Msun, which allows us to study the
  low-mass galaxy population. The 20 deg$^2$ main survey area is
  centred on fields with exceptional multi-wavelength ancillary data,
  with photometry ranging from optical through far-infrared
  wavelengths, supplemented with multiple spectroscopic campaigns. We
  describe here the survey design and the key science goals. We also
  show first results from the Early Science observations, including
  kinematic modelling of individual sources, along with the redshift,
  \HI, and stellar mass ranges of the sample to date.}


\titlerunning{The MIGHTEE-HI project}
\authorrunning{N. Maddox et al.}
\maketitle


\section{Introduction}

The neutral hydrogen (\HI) content of galaxies serves as the raw
material for the build-up of stellar mass. It is therefore essential
that this component be incorporated into our multi-wavelength census
of galaxy evolution. Unlike the stellar mass, the \HI\ content of
galaxies is a non-monotonic quantity, with gas being accreted, expelled,
re-accreted, stored, and consumed. As such, we would like to observe the \HI\ in
galaxies as a function of cosmic time. Until recently, this was hindered by
technological limitations. Radio telescopes with small bandwidth
restricted observations to narrow redshift ranges. The intrinsic
faintness of the \HI\ signal coupled with
the limited sensitivity of existing facilities meant that regular observations
extending beyond the local Universe were out of reach. 

Several large-area surveys have been undertaken to detect \HI\ in the
low-redshift ($z<0.1$) Universe. In the Southern Hemisphere, the \HI\
Parkes All-Sky Survey (HIPASS; \citealt{Barnes2001}), complemented by
the Northern HIPASS extension (NHICAT; \citealt{Wong2006}) and the
Effelsberg Bonn \HI\ Survey (EBHIS; \citealt{Kerp2011}) in the north,
cover the entire sky. The Arecibo Legacy Fast ALFA Survey (ALFALFA; 
\citealt{Giovanelli2005}) provides improved spatial 
resolution and sensitivity over 6900 deg$^2$. These large-area surveys
were performed blindly over their 
limited bandwidth and are invaluable as they provide the benchmark
for the $z=0$ cosmic \HI\ density \citep{Jones2018a}. 

At higher redshifts, targeted surveys such as the Blind Ultra-Deep
\HI\ Environmental Survey (BUDHIES; \citealt{Verheijen2007};
\citealt{Jaffe2013}) and the HIGHz survey \citep{Catinella2015} spent many
hundreds of hours of observations to detect individual galaxies at
$z\sim 0.2$. While providing the first look at \HI\ in galaxies at
cosmological distances, the targets were chosen in advance based on
their optical properties, thus they are
not necessarily representative of the general galaxy population.

Technological advances, aided by increased computing capacity, are
enabling the next generation of radio telescopes to explore previously
inaccessible parameter space, undertaking blind \HI\ surveys spanning
substantial redshift ranges. The number of new \HI
-focused surveys indicates the importance of these observations for
our understanding of galaxy evolution. The recently upgraded
Karl G. Jansky Very Large Array (VLA) has the bandwidth in the L-band to observe the
\HI\ spectral line simultaneously from redshift zero to $z\sim 0.42$.
The COSMOS \HI\ Large Extragalactic Survey
(CHILES) is exploiting this expanded frequency range to be the first
next-generation deep \HI\ survey, having already acquired the highest
redshift direct \HI\ detection to date, at $z=0.376$ \citep{Fernandez2016}. At lower
frequencies, the upgraded Giant Metrewave Radio Telescope (uGMRT) has
recently produced statistical \HI\ detections at $z\sim 0.3$
\citep{Bera2019} and has the frequency coverage to detect \HI\ to
$z\sim 1$ via statistical methods.

With a much-expanded field of view (FoV) relative to its predecessor,
the APERture Tile in Focus 
(Apertif; \citealt{Verheijen2008}; \citealt{Oosterloo2009})
phased array feed system now deployed on the Westerbork 
Synthesis Radio Telescope (WSRT) upgrades the existing facility to a
wide-field survey instrument. The imaging surveys underway include a
wide and a medium-deep tier, designed to cover 3500 and 350 deg$^2$,
respectively, in the full survey plan, with spectral coverage for \HI\ 
to $z=0.26$ (\citealt{Adams2019}; Hess et al. in prep.).

The next generation of newly built facilities, employing new
technologies, further open parameter space that has eluded existing
instrumentation. The Australian Square Kilometre Array Pathfinder
(ASKAP; \citealt{Johnston2008}) employs a new Phased Array Feed (PAF)
receiver design, similar in concept to Apertif, to 
provide 30 deg$^2$ of instantaneous coverage. This large FoV will enable 
the Widefield ASKAP L-band Legacy All-sky Blind surveY (WALLABY;
\citealt{Koribalski2020}) to 
observe the entire Southern Sky, providing increased 
redshift sensitivity and much-improved spatial resolution over the
existing HIPASS data. The Deep Investigation of Neutral Gas Origins
(DINGO; \citealt{Meyer2009}) survey provides the corresponding deep
and ultra-deep tiers, spending 
several thousands of hours integrating to detect \HI\ to $z\sim 0.4$
over tens of square degrees.

\subsection{MeerKAT}\label{subsec:meerkat}

Observing \HI\ over large areas covering cosmological ranges in
redshift is one of the key science goals of the Square Kilometre Array
(SKA; \citealt{Braun2015}), and is within the capabilities of the SKA
precursor facility MeerKAT \citep{Jonas2016}. The 64-dish array,
equipped with L-band ($900<\nu <1670$ MHz), UHF-band ($580<\nu
<1015$ MHz) and higher frequency S-band ($1750<\nu <3500$ MHz)
receivers, is the most sensitive centimetre-wavelength interferometer
in the Southern Hemisphere, only to be surpassed by the SKA.
Although MeerKAT employs a single-pixel feed rather than a PAF,
the 13.5-m offset Gregorian dishes provide a $>$1 deg$^2$
field-of-view (FoV) at 1420 MHz (i.e. $z=0$ for \HI\ observations),
$\sim 4\times$ larger than the FoV for the VLA.
The system equivalent flux density (SEFD) of the MeerKAT dish
operating at L-band has been measured and found to be similar to the
larger (25-m) VLA dishes at similar frequencies, due to MeerKAT's much
lower system temperature \citep{Jonas2016}. The factor of $\sim 2$ 
increase in number of dishes, 64 for MeerKAT compared to 27 for VLA,
thus reduces observing time to achieve a given flux limit by a factor of four.
Combining the increased sensitivity and larger FoV, the survey speed
of MeerKAT is approximately 16 times that of the VLA at L-band.
The distribution of antennas simultaneously
provides good angular resolution ($\sim 10$ arcsec) while 
retaining sensitivity to low surface brightness and extended
features. A stunning example of the performance of the array and
quality of the data is the image of the Galactic Centre in radio
continuum \citep{Heywood2019}. 

A combination of depth and area coverage are essential to fully
parametrise the \HI\ Universe. Sensitivity is required to detect the
intrinsically faint \HI\ signals, both from nearby low-mass objects,
and more distant higher-mass galaxies. As \HI\ is known to be 
sensitive to galactic environments, large volumes are required to fully sample
a range of environments at a given redshift, including galaxies in the
field. Therefore, MeerKAT is the ideal instrument with which to
undertake a deep, extragalactic, large-area (tens--thousands of square
degrees), blind \HI\ survey. 

MeerKAT will undertake eight Large Survey Projects (LSPs) before the
SKA begins operations. Several LSPs have significant
  \HI\ focus, indicating the advances in this field MeerKAT will
  bring. The MeerKAT \HI\ Observations of Nearby Galactic Objects;
  Observing Southern Emitters survey (MHONGOOSE; \citealt{deBlok2016})
  focuses on nearby, 
  well-resolved galaxies, detecting the neutral gas to low column densities
  that probe the connection between galaxies and the intergalactic
  medium. The MeerKAT Fornax Survey (MFS; \citealt{Serra2016}) is
  centred on the nearby Fornax cluster, investigating the \HI\ in galaxies
within this dense and dynamic environment. Looking
At the Distant Universe with the MeerKAT Array (LADUMA;
\citealt{Blyth2016}) will spend thousands of hours of integration on a
single pointing to give the deepest view of the \HI\ universe,
extending to $z=1.4$. Also extending to high redshifts is the MeerKAT
Absorption Line Survey (MALS; \citealt{Gupta2016}), looking instead to
detect \HI\ in absorption towards background radio continuum sources.

Here we describe the MeerKAT International GigaHertz Tiered
Extragalactic Exploration survey (MIGHTEE; \citealt{Jarvis2016}) LSP, 
focusing on the \HI\ science enabled within the survey.
In Sect.~\ref{sec:survey}, we describe the details of the survey, along
with predicted galaxy number counts. In Sect.~\ref{sec:science}, we outline
the key science goals of the project. In Sect.~\ref{sec:progress}, we
outline the data products that will be delivered and provide
illustrative examples. 
Section~\ref{sec:summary} gives a summary of the work. Concordance
cosmology with $H_{0} = 67.4$ km s$^{-1}$ Mpc$^{-1}$ (thus $h\equiv 
H_{0}$/[100 km s$^{-1}$ Mpc$^{-1}$]$=0.674$), $\Omega_{\rm m} = 0.315$,
$\Omega_{\Lambda} = 0.685$ is assumed \citep{Planck2018}, and AB
magnitudes are used throughout unless otherwise stated.

\section{Survey description}\label{sec:survey}

The parent survey for this project is the MeerKAT International
GigaHertz Tiered Extragalactic Exploration survey (MIGHTEE;
\citealt{Jarvis2016}). MIGHTEE is one of eight guaranteed-time LSPs
to be undertaken with MeerKAT. The survey will cover 20
deg$^2$ divided over four fields (see Table~\ref{tab:fields}). The
nominal integration time per MeerKAT pointing is $\sim$16 hours, but
due to overlap within the mosaic pointings, the effective integration
time for points within the mosaic increases to $\sim$23 hours. The
continuum confusion floor with a 6--8 arcsec synthesised beam at $\sim
1000$ MHz is reached with integrations longer than this.
Full-depth continuum imaging has yielded a confusion-limited image
with a thermal noise floor of $\sim$2\,\uJy\ beam$^{-1}$, in line
with the survey design predictions. The actual observing strategy may evolve
from this nominal plan as the survey progresses.

The MIGHTEE collaboration is organised into working groups that
focus on radio continuum, \HI\ in emission, \HI\ in absorption, and
polarisation. The working groups closely collaborate as there is
substantial overlap between the data products and respective science cases.

\subsection{MIGHTEE-HI}\label{subsec:mighteehi}

The MeerKAT data are collected in spectral line mode, so the MIGHTEE
survey is a fully commensal continuum, polarisation, and spectral line 
project. The key metrics for the \HI\ science to be undertaken with
MIGHTEE, hereafter designated as MIGHTEE-HI, are listed in
Table~\ref{tab:params}.

To maximise scientific return from MeerKAT observations, the MeerKAT
Fornax Survey \citep{Serra2016} and 
MIGHTEE have agreed that the volume behind the Fornax cluster will be
analysed as part of MIGHTEE. The division between the two surveys is
set to be at 110 Mpc, corresponding to $z=0.025$, or $\nu = $1385 MHz,
and increases the effective area of MIGHTEE-HI to 32 deg$^2$ (see
Table~\ref{tab:fields}). The MFS mosaic pattern is slightly more closely packed than
for the MIGHTEE survey, with an effective integration time of 25 hours
per sky location, thus the per-channel noise level for MFS and
MIGHTEE-HI will be comparable.

MIGHTEE-HI is strongly complementary with other next-generation
emission-line surveys. In 
particular, the planned DINGO Deep tier will cover 150 deg$^2$ over
$0<z<0.26$ while the Ultradeep tier extends to $0.1<z<0.43$ over 60
deg$^2$. MIGHTEE-HI covers half the area of the Ultradeep tier, but
with greater sensitivity, better spatial resolution and higher maximum
redshift. The Apertif medium-deep survey is comparable to the DINGO Deep
tier, focusing on the Northern Hemisphere.

The ultra-deep CHILES project has 1000 hours of integration of a
single pointing with the VLA in B configuration, and has better
spatial resolution ($\sim 5$ arcsec at 1420 MHz) than MeerKAT, but
lower sensitivity to extended structures (Jarvis et al. in prep.).
The expected sensitivity of CHILES at 1420 MHz, $\sim$50\,\uJy\
  beam$^{-1}$ per 31 kHz channel \citep{Fernandez2016}, is a factor of
  two deeper than 
  the $\sim$100\,\uJy\ beam$^{-1}$ sensitivity per 26kHz channel for
  MIGHTEE-HI.  However, the area covered is only 0.25 deg$^2$,
significantly smaller than the  
32 deg$^2$ contained within MIGHTEE. The MeerKAT LSP LADUMA
is conceptually similar to CHILES, spending
thousands of hours of integration on a single pointing, but employs
both the L- and UHF-bands to extend the sensitivity to \HI\ to
$z=1.4$. LADUMA takes advantage of MeerKAT's increasing FoV with
redshift to encompass more volume, ranging from $\sim$1 deg$^2$ at
$z=0$ to $>$5 deg$^2$ at the highest redshift. MIGHTEE and LADUMA
are complementary in area and depth, as illustrated in
Sect.~\ref{subsec:himf}. 

While MIGHTEE-HI, and indeed any of the deep or ultra-deep surveys,
cannot compete in area covered with the wide-area surveys such as
WALLABY, they do contain substantial cosmological volumes. The 
volume contained within 32 deg$^2$ and $0<z<0.58$ is 0.033 Gpc$^3$,
almost three times that of the HIPASS survey, which surveyed $\delta
< 2$ and $z<0.0423$. MIGHTEE-HI also extends 2.5 Gyr of
cosmological lookback time farther than WALLABY.

In full survey mode, MIGHTEE is being observed with a spectral resolution of 26 kHz,
corresponding to 5.5\,\kms\ channels at $z=0$.
The MeerKAT 64-dish array consists of a compact core, with
70~per~cent of the dishes contained within a diameter of approximately
1.5 km, and the remaining dishes distributed out to a maximum
baseline of 7.7 km. The array configuration produces optimal
sensitivity behaviour between resolutions of 8--80 arcsec
at 1420 MHz \citep{Jonas2016}. Spectral line imaging based on
MIGHTEE observations shows that using a Briggs  
robustness parameter of 0.5 produces a nominal resolution of $\sim$ 12
arcsec.

The raw visibility data is converted to Measurement Set (MS) format,
and is transferred from the MeerKAT archive to the Ilifu
cluster\footnote{\url{http://www.ilifu.ac.za/}} hosted at the
University of Cape Town. Prior to calibration, we flag
three frequency ranges corresponding to bands of persistent radio
frequency interference (RFI), at 933--960\,MHz, 1163--1310\,MHz, and
1524--1630\,MHz. Once flagged, the data is further split
into multi measurement sets (MMSs) corresponding to 50\,MHz
windows. The full-Stokes \textit{a priori} or cross calibration is
done using the IDIA \texttt{processMeerKAT}
pipeline\footnote{\url{https://idia-pipelines.github.io}} (Collier et
al., in prep.); each spectral window is
processed concurrently. Bandpass calibration is done using either
J1939-6342 or J0408-6546. The secondary calibrator is used to solve
for antenna-dependent gains, leakages and antenna XY phase. At the end
of cross-calibration, the MMSs are reassembled and self-calibration is
done on the downsampled data (averaged by a factor of four). Three
loops of phase-only gain calibration is done for each observation,
and the resulting solutions are applied to the full resolution data
using linear interpolation in time and frequency.  

The first step in continuum subtraction involves the subtraction of
the final model generated by the self-calibration process. Polynomial
fits to the visibility data are subtracted in the second step, for
each spectral window. Spectral line imaging is performed on the
resulting data. A final step of median filtering is done on the output
image cubes, in order to reduce any remaining artefacts due to
direction-dependent errors. See Frank et al. (in prep.) for a
full description of the data reduction procedure.

For an effective integration time of 23.4 hours, 26 kHz
channels, and the system equivalent flux density (SEFD) from the
MeerKAT public
documentation\footnote{\url{http://public.ska.ac.za/meerkat/meerkat-schedule}\\
  Version release 2016-10-10}, we expect to reach a sensitivity of
$\sim$100\,\uJy\ beam$^{-1}$ per channel at 1420 MHz. At an angular
resolution of 12 (30) arcsec, the corresponding 
3-$\sigma$ column density sensitivity limits for MIGHTEE-HI are 
$1.3\times 10^{19}$ ($2.0\times10^{18}$)\,cm$^{-2}$\ at spectral resolution
of 26 kHz, and $2.6\times10^{19}$ ($4.0\times10^{18}$)\,cm$^{-2}$\ over 104
kHz, or 22\,\kms, channels. The bulk of the \HI\ mass
  associated with the main disk in galaxies is typically at
  $>1.25\times10^{20}$\,cm$^{-2}$, well within the sensitivity of the
  observations. The different science cases outlined in
Sect.~\ref{sec:science} benefit from data products at optimised
spatial and spectral resolutions. 

\begin{table}
\centering
\caption{List of the four fields within the MIGHTEE footprint. The fifth field
  is the MeerKAT Fornax Survey field, providing an additional 12
  deg$^2$ of area beyond $z=0.025$. The areas are computed at 1420 MHz,
  but become larger 
  as the MeerKAT FoV increases with decreasing
  frequency. $^a$A single pointing in the centre of ECDFS is
  covered by the LADUMA survey, surrounded by the MIGHTEE
  pointings. See \citet{Jarvis2016} for details.}
\label{tab:fields}
\begin{tabular}{|c|c|c|}\hline
  Field & Area (deg$^2$) & Centre \\
  Name & Full Survey & Coordinates \\ \hline
  COSMOS & 2 & 10h01m, +02d12m \\ 
  XMMLSS & 8 & 02h20m, -04d50m \\ 
  ECDFS$^a$ & 8 & 03h32m, -28d00m \\ 
  ELAIS-S & 2 & 00h40m, -44d00m \\
  MFS & 12 & 03h38m, -35d27m \\ \hline 
\end{tabular}
\end{table}

\begin{table}
\centering
\caption{Key parameters of the MIGHTEE-HI observations}
\label{tab:params}
\begin{tabular}{|l|c|}\hline
  Area covered & 32 deg$^2$ \\
  Frequency range& 900--1420 MHz \\
  Redshift range for \HI & $0<z<0.58$ \\
  Nominal angular resolution & 12 arcsec \\
  Velocity resolution & 5.5\,\kms\ (1420\,MHz) \\ 
  Per channel flux sensitivity & 100\,\uJy\ beam$^{-1}$ (1420\,MHz) \\ 
  Column density sensitivity & $2.6\times 10^{19}$\,cm$^{-2}$ \\
   (22\,\kms) & \\ \hline
  
\end{tabular}
\end{table}

\subsection{Expected number counts}\label{subsec:sims}

We can predict the approximate number of galaxies we expect to detect
within MIGHTEE-HI, and their mass and redshift distribution, from
upgraded simulations based on those from \citet{Maddox2016}. Briefly,
we assume the \HI\ mass function (HIMF) from \citet{Jones2018a} for the full ALFALFA
survey, and we conservatively do not allow this to evolve with redshift. For a
representative galaxy profile 150\,\kms\ wide, the per-channel sensitivity of
$\sigma =100$\,\uJy\ beam$^{-1}$ translates to a 5-$\sigma$ detection
level of 0.075 Jy \kms\ beam$^{-1}$. We note that this value is too large
for low-mass galaxies, as they have \HI\ profile widths substantially
narrower than 150\,\kms, thus the number of these galaxies 
predicted to be detected by the simulations is an
underestimate. A more careful approach to the detection predictions
for low-mass galaxies is taken in Sect.~\ref{subsec:lowmass}. We also
note that the simulations here are strictly flux-limited, rather than
surface-brightness limited. For example, a
resolved galaxy can be below the per-beam flux limit, but if it is spatially
resolved, it can still be detected. As described in
Sect.~\ref{subsec:resolved}, the majority of \HI -detected galaxies
beyond the local Universe 
will not be well-resolved, so this issue is minor and the results of
the simulations are still illustrative. The 
effects of cosmic variance are ignored.

\begin{figure}
\includegraphics[width=\columnwidth]{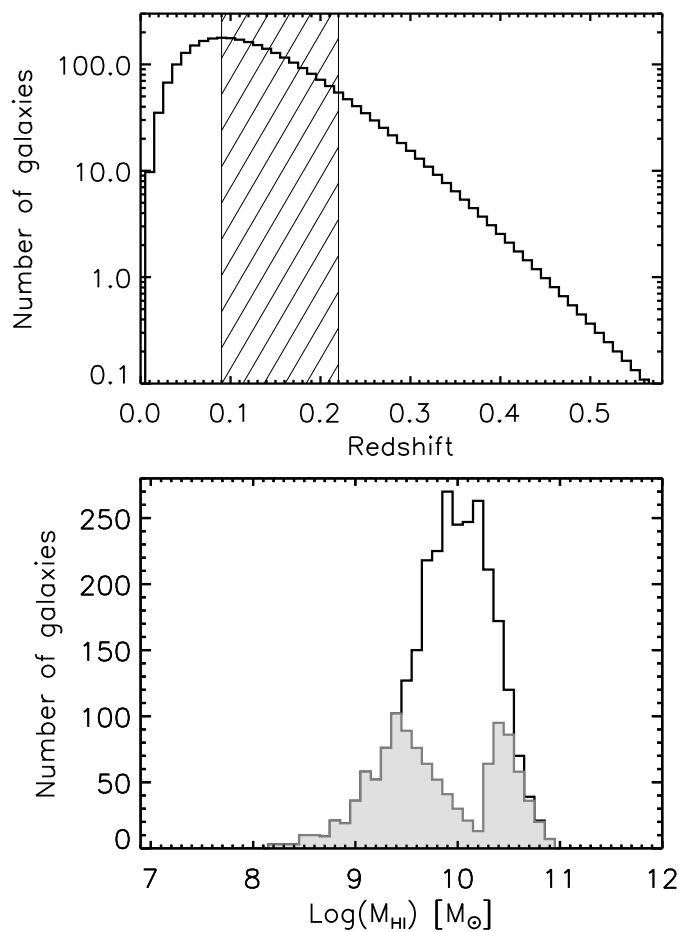}
\caption{Conservative number of galaxies expected from the full 32
  deg$^2$ of MIGHTEE-HI area, derived from simulations. (Top) Expected
  number of galaxies from 32 deg$^2$ as a function of redshift. The
  redshift range most affected by RFI is hatched out. (Bottom)
  Expected number of galaxies from 32 deg$^2$ as a function of
  \MHI. The solid-line histogram is the distribution covering the full
  redshift range, whereas the light filled histogram represents the
  worst-case of fully excluding the redshift range hatched out in the
  top panel.}
\label{fig:sims}
\end{figure}

In the top panel of Fig.~\ref{fig:sims}, the predicted number of
detected galaxies as a function of redshift for 32 deg$^2$ is shown, with
galaxies directly detected to $z\sim 0.4$. While the MeerKAT L-band
spans 900--1670 MHz, resulting in sensitivity to \HI\ 
over $0<z<0.58$, radio frequency interference (RFI) from global
navigation satellite systems (GNSS) at 1164--1300 MHz,
which corresponds to $0.09<z<0.22$ for \HI, affects the data
quality in that frequency range. RFI from GNSS is seen at all
radio observatories, and thus affects all \HI\ observations at those
frequencies. The hatched region in Fig.~\ref{fig:sims} corresponds
to the frequency range most affected by RFI. In the ideal case with no
RFI, we expect to detect $\sim 1750$ galaxies to be individually
detected in the main 20 deg$^2$ MIGHTEE-HI area, with an extra $>1000$
galaxies in the 12 deg$^2$ MeerKAT Fornax Survey area. In the most pessimistic case,
excluding the frequency range affected by RFI entirely, these numbers
drop to lower limits of 730 in the main area and 420 in the Fornax area.
In practice, there are narrow frequency ranges
that are less affected than others and can be salvaged. RFI is also
less prominent at increasingly long baselines, so removing data from
short baselines, while accepting the associated loss in sensitivity,
is another method for alleviating the issue. See Figure 7 in
\citet{Mauch2020} for a MeerKAT L-band spectrum illustrating the RFI environment,
and the benefit gained by removing short baselines.

The bottom panel of Fig.~\ref{fig:sims} shows the expected distribution 
of \HI\ mass (\MHI ) of galaxies within the 32 deg$^2$ MIGHTEE-HI
area. The solid-lined 
histogram includes galaxies across the full redshift range, whereas
the light-filled histogram corresponds to the worst-case scenario of
assuming no useful data can be recovered from the redshift range
$0.09<z<0.22$, and therefore represents a lower limit. Advancing RFI
identification and mitigation techniques currently underway will clearly have a
large beneficial effect on \HI\ observations undertaken at these
frequencies, including for future SKA surveys.

\subsection{Ancillary data}\label{subsec:ancdata}

\begin{figure}
\includegraphics[width=\columnwidth]{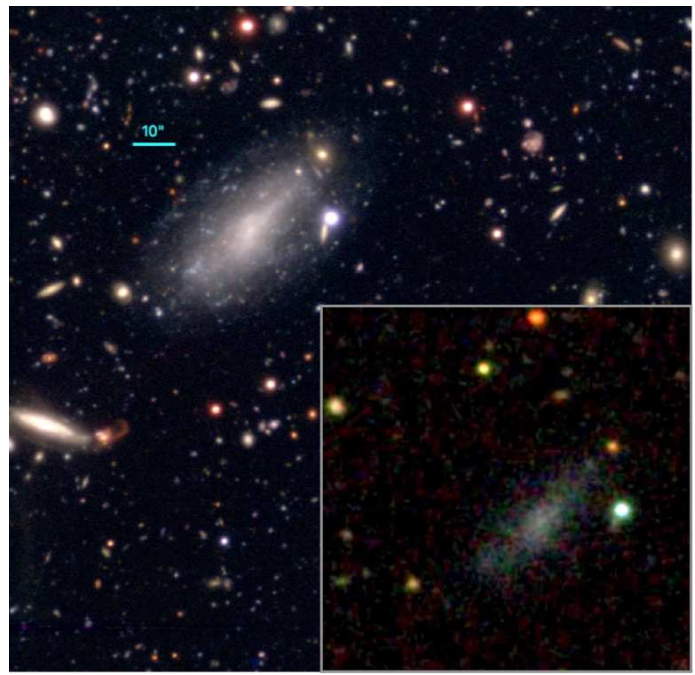}
\caption{HSC $gri$ false-colour image of a galaxy at $z=0.0055$ detected within MIGHTEE-HI, with
  the SDSS $gri$ image on the same size scale inset. The depth and image quality
  of the HSC images reveal structural details and low surface
  brightness features not visible in shallower imaging.}
\label{fig:composite}
\end{figure}

The four fields central to MIGHTEE listed in Table~\ref{tab:fields}
are some of the most well-studied 
extragalactic fields accessible from the Southern Hemisphere and are
covered by various surveys ranging from the X-ray through far-infrared
wavelengths. While long integrations are required to detect the most
distant objects, the resulting depth is also valuable for low-redshift
galaxies, revealing details not visible in lower quality imaging, as
illustrated in Fig.~\ref{fig:composite}. The redshift of the galaxy,
$z=0.0055$, is known from the \HI\ detection, and would be difficult
to obtain from optical spectroscopy due to its intrinsic faintness and
low surface brightness.

A tabulated description of the data sets available in
XMMLSS, ECDFS and ELAIS-S1, across all wavelengths, can be found in
Table~1 of \cite{Chen2018}, and the data available in the COSMOS field
is generally greater in
extent\footnote{\url{http://cosmos.astro.caltech.edu}}. The key data
sets for the \HI\ science described here are the optical photometry 
from Canada-France-Hawaii-Telescope Legacy Survey (CFHTLS) and
HyperSuprimeCam \citep[HSC; ][]{Aihara2018,Aihara2019}. The
near-infrared data are from the VISTA Deep Extragalactic Observations
\citep[VIDEO; ][]{Jarvis2013} and the VISTA Extragalactic Infrared
Legacy Survey \citep[VEILS; ][]{Honig2017}, which together cover the
MIGHTEE XMMLSS, ECDFS and ELAIS-S1 fields, while UltraVISTA
\citep{McCracken2012} covers the COSMOS field. These data have been
matched in pixel-space and catalogues have been made, providing
photometry based on a range of apertures
as described in \citet{Bowler2020}, alongside photometric redshift
catalogues \citep{Adams2020}. A source catalogue 
derived using ProFound \citep{Robotham2018} to obtain
accurate photometry for the low redshift, large angular size galaxies detected
in \HI\ within the deep imaging data will also be available (Davies et al. in
prep.; Adams et al. in prep.).

The fields are also being covered by past, current and future spectroscopic
campaigns. Past surveys include the VIMOS VLT Deep Survey (VVDS;
\citealt{LeFevre2013}), VANDELS \citep{McLure2018,Pentericci2018},
Z-COSMOS \citep{Lilly2009}, the Sloan Digital Sky Survey Data Release
12 (SDSS-DR12; \citealt{Alam2015}), 3D-HST \citep{Skelton2014,Momcheva2016},
the PRIsm MUlti-object Survey (PRIMUS; \citealt{Coil2011,Cool2013}),
DEIMOS-10K \citep{Hasinger2018} and the Fiber Multi-object
Spectrograph (FMOS) COSMOS Survey \citep{Silverman2015}. Many of these
large, extragalactic spectroscopic surveys target the high-redshift
Universe, beyond the reach of \HI\ surveys.

Three programmes, one ongoing and two soon to start, seek to
  collect more complete data over the redshift range $0<z<1$. 
The Deep Extragalactic VIsible Legacy Survey (DEVILS,
\citealt{Davies2018}) is an ongoing survey being carried out with the
multi-object spectrograph on the Anglo-Australian Telescope (AAT),
within the COSMOS, XMMLSS and ECDFS fields.
The target redshift range is $0.3<z<1.0$, aiming for very high
(95~per~cent) completeness to $Y<21.2$, and will be a key data set for
MIGHTEE-HI science, in particular for stacking, as outlined in
Sect.~\ref{subsec:stacking}. In the coming years, DEVILS will be
complemented by surveys with the Multi-Object Optical Near-Infrared Spectrograph
\citep[MOONS; ][]{Cirasuolo2014} and the Wide-Area VISTA Extragalactic
Survey \citep[WAVES; ][]{Driver2019} with the 4-m Optical
Spectroscopic Telescope \citep[4MOST; ][]{deJong2019}. WAVES aims for
90~per~cent completeness to $z<0.8$ and $Z<21.25$ within the 4 primary MIGHTEE
fields, probing below the characteristic stellar mass over the full
redshift range. The high completeness and stellar mass-based selection
will alleviate biases from galaxy type and small-scale galaxy
environment. We also note that these deep surveys are complemented by
spectroscopy from the SDSS in both COSMOS and XMMLSS, and the all-sky
2MASS Redshift Survey (2MRS; \citealt{Huchra2012}), which provide
useful spectroscopic information on the bright, low-redshift galaxy population. 

These fields will also be covered by the Vera C. Rubin Observatory as
part of the deep drilling fields \citep{Jones2013}, providing
extremely deep visible wavelength imaging, and the wide-
and deep-survey to be carried out with the {\sc
  Euclid} satellite \citep{Laureijs2010}, providing deep and
high-resolution near-infrared imaging.

The MFS area has recently been observed with the VLT Survey Telescope
(VST), providing imaging in the SDSS $u, g,
r$ and $i$ bands, with surface brightness sensitivity extending to
27--28 mag arcsec$^{-2}$ \citep{Venhola2018}. Spectroscopy over the
full field, including existing and new observations, has been compiled
by \citet{Maddox2019}. Imaging in 5 optical bands from the Dark Energy Survey (DES;
\citealt{Flaugher2005}) also covers the MFS area. 

\section{Key science goals}\label{sec:science}

There are a number of key science goals to be addressed with the
MIGHTEE-\HI\ data, which take advantage of the unique combination of
depth, area and available ancillary data. Below is a list of the primary science 
topics to be addressed. 

\subsection{The \HI\ mass function}\label{subsec:himf}

A headline goal of the MIGHTEE-\HI\ project is the
determination of the HIMF, as a function of
redshift. This fundamental quantity describes the distribution of \HI\
within galaxies. The most recent constraints on the $z\sim 0$ HIMF parameters,
including the low-mass slope, the characteristic mass, and the
normalisation, come from the ALFALFA survey \citep{Jones2018a}. This
work, covering 6900 deg$^2$ and containing more than 25,000 sources,
illustrates the effects of large-scale structure and sample
variance on the HIMF determination, particularly on the low-mass
slope.

Instead of wide-area and shallow, the Arecibo Ultra-Deep Survey (AUDS;
\citealt{Hoppmann2015}) used deep integrations with the Arecibo
telescope over a smaller area to detect 102 galaxies over 1.35 deg$^2$.
This extended the HIMF parametrisation to the mean redshift of their
sample at $z=0.065$.  

\begin{figure}
\includegraphics[width=\columnwidth]{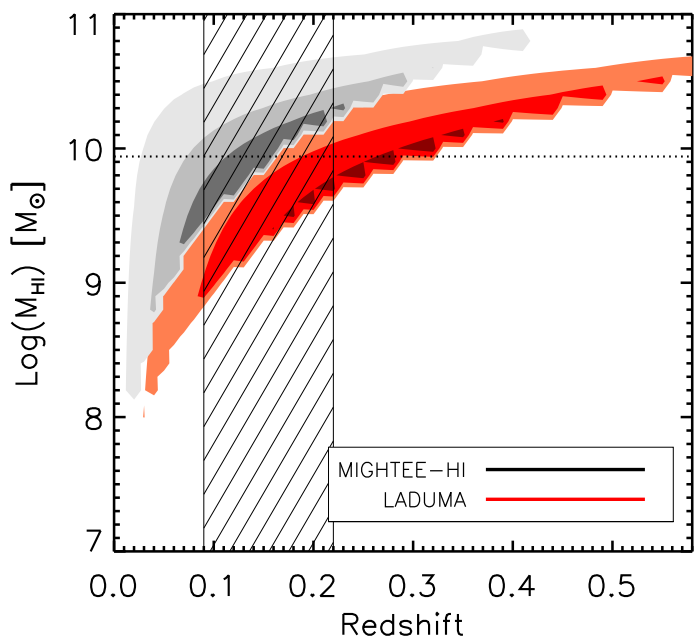}
\caption{\MHI\ -- redshift parameter space, showing where
  MIGHTEE-HI and LADUMA are sensitive to galaxies, divided into cells of size
  $\Delta z$ = 0.01 and $\Delta$ log(\MHI ) = 0.1. Light, medium, and dark
  grey shading indicates where at least one, 10, and 25 galaxies are
  expected per cell. Light, medium, and dark red shading shows where at
  least one, four, and eight galaxies are expected per cell. The horizontal
  dotted line indicates the $z=0$ characteristic \HI\ mass from the HIMF.
   The hatched region corresponds to the redshift range $0.09<z<0.22$
   most affected by RFI.}
\label{fig:Mig_Lad}
\end{figure}

While predicted to contain an order of magnitude fewer sources than ALFALFA,
MIGHTEE-HI extends to higher redshifts, affording us a first
opportunity to investigate the HIMF as a function of redshift to
$z=0.1$ and beyond. LADUMA is also aiming to constrain the HIMF
at $z>0$. In Fig.~\ref{fig:Mig_Lad}, the parameter space is divided
into cells $\Delta z$ = 0.01 and $\Delta$log(\MHI ) = 0.1, and shows where
MIGHTEE-HI and LADUMA are expected to directly detect galaxies. For MIGHTEE-HI,
the light grey shaded region indicates cells within which 
at least one galaxy is predicted to be detected by the simulations,
medium grey shading is 10 galaxies per cell, and dark grey is 25
galaxies per cell. In reality, the shaded regions do not have sharp
edges, primarily due to inclination effects that have not been
accounted for, but the simulations are still useful for illustration.
For LADUMA, the light, medium and dark red shaded
regions indicate cells with at least 1, 4 and 8 galaxies predicted by
the simulations. LADUMA in fact employs both the L- and UHF-bands on
MeerKAT, extending sensitivity to \HI\ to $z=1.4$, but we show only
the redshift range covered by the L-band here.

From Fig.~\ref{fig:Mig_Lad}, MIGHTEE-\HI\ and LADUMA probe
complementary parameter space in \HI\ mass, thus the combination of the two
surveys will enable parametrisation of the HIMF to $z\sim 0.3$ using
direct detections, and higher redshifts using statistical techniques.
In particular, the characteristic \HI\ mass is well covered by
MIGHTEE-HI at $z<0.15$. Below the formal detection threshold, a
statistical approach to 
determining the HIMF using MIGHTEE-HI data can also be employed, as in
\citet{Pan2020}. This technique will exploit the relatively large
volume probed by MIGHTEE-HI at $0.2<z<0.6$.

Integration of the HIMF provides a measure of the cosmic \HI\ gas
density, \omegahi. A compilation from \citet{Rhee2018} (see their
Figure 14) shows an increasing number of measurements of \omegahi\ at
low and intermediate redshifts. However, care must be used when
interpreting this collection of results. Density measurements from directly
detected \HI\ extend to $z<0.127$, with the highest redshift
determination from AUDS \citep{Hoppmann2015}. Beyond that,
to $z\sim 0.4$, results are derived from statistical methods,
primarily \HI\ stacking experiments. At $z>0.4$, it is \HI\ in
absorption towards background quasars being measured. With MIGHTEE-HI,
coupled with the LADUMA survey, \omegahi\ can be measured from
directly detected \HI\ to at least $z=0.3$, and possibly beyond. Statistical
methods can be used to $z=0.58$, which overlaps in redshift with the
\omegahi\ probes derived from \HI\ absorption. We will thus, for the
first time, have independent measures of \omegahi, one from \HI\
emission, and one from \HI\ absorption, at overlapping redshifts. 

\subsection{Kinematics from spatially resolved \HI}\label{subsec:resolved}

Resolved galaxies in the local Universe have been 
extensively studied in the past. The main scientific topics 
based on resolved \HI\ can be summarised in two broad  
categories. First, the study of galaxy gas accretion and depletion,
which includes minor mergers (e.g. \citealp{DiTeodoro2014}),  
hot-mode accretion (e.g. \citealp{Heald2011,Marasco2019}), ram-pressure 
stripping (e.g. \citealp{Jaffe2015,Yoon2017}), outflows and feedback 
(e.g. \citealp{Bagetakos2011, Maccagni2017}). This is further expanded
in Sect.~\ref{subsec:accretion}. Second, the study of 
galaxy structure and kinematics, which includes warps and lopsidedness 
(e.g. \citealp{vdKruit2007}), rotation curves and dark matter 
 (e.g. \citealp{Ponomareva2016,Aniyan2018}), angular momentum 
(e.g. \citealp{Obreschkow2016,Posti2018}), non-circular motions 
(e.g. \citealp{Shetty2008}). Though these topics are scientifically vast, 
almost all take advantage of kinematic modelling.

\begin{figure}
\includegraphics[width=\columnwidth]{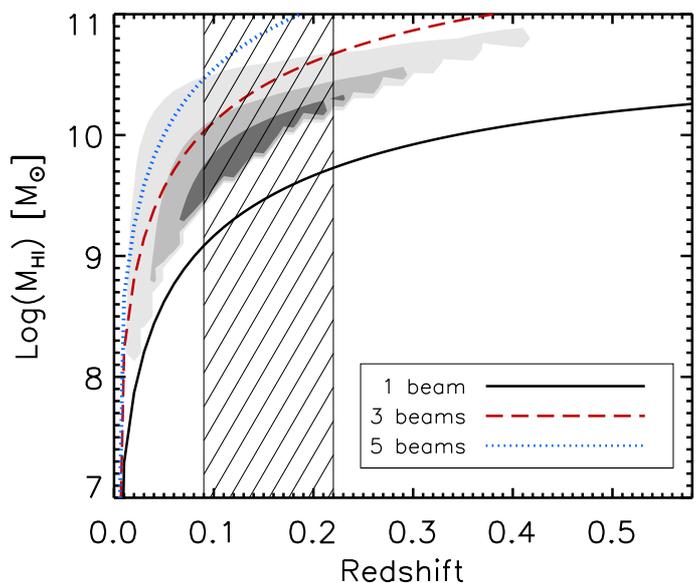}
\caption{\MHI\ -- redshift parameter space, showing where
  MIGHTEE-HI is sensitive to galaxies, divided into cells of size
  $\Delta z$ = 0.01 and $\Delta$ log(\MHI ) = 0.1. Light, medium, and dark
  grey shading indicates where at least one, 10, and 25 galaxies are
  expected per cell. The black, red, and blue curves correspond to the
  \MHI\ of galaxies as a function of redshift, for three 
fixed diameters, using the \MHI\ -- \DHI\ relation from
\citet{Wang2016}. The black line shows the 12 arcsec MeerKAT beam,
indicating galaxies below this mass will be unresolved. The red
dashed line is for marginally resolved sources and the blue dashed
line is for well-resolved sources with five beams across. $0.09<z<0.22$ is
worst-afftected by RFI, and the size and shape of the synthesised beam is strongly
affected in this region.}
\label{fig:resolved}
\end{figure}

Figure~\ref{fig:resolved} illustrates the mass and redshift range of
MIGHTEE-HI galaxies that will be spatially resolved for kinematic modelling. 
The grey shading is the same as in Fig.~\ref{fig:Mig_Lad}. 
The well-established relation between the \HI\ mass of a galaxy and
its diameter, \DHI, as recently investigated by \citet{Wang2016}, can
be used to predict how many galaxies within MIGHTEE-HI will be
resolved. Assuming a MeerKAT beam major axis of 12 arcsec, this is
converted to a linear size in kiloparsecs as a function of redshift,
and then converted to an equivalent \MHI. The \MHI\ corresponding to
the MeerKAT beam size is shown as the black curve in
Fig.~\ref{fig:resolved}. 

Kinematic modelling requires at least three resolution elements (with
the signal-to-noise ratio of the data larger than 2-3) across
the major axis of a galaxy in order to constrain the dynamics of the
gas \citep{DiTeodoro2015}. The \MHI\ corresponding to three times the
MeerKAT beam is shown as the red dashed line in
Fig.~\ref{fig:resolved}, and we refer to this as minimally resolved. The blue dotted
line corresponds to 5 MeerKAT beams across a galaxy, which is
generally the requirement for a galaxy to be considered
well-resolved. We expect $\sim$ 200 galaxies to be minimally
resolved out to $z<0.09$, with up to 50 being well resolved. No
galaxies will be even minimally resolved at $z>0.22$ using the
expected 12 arcsec beam, but different weightings could be used to
increase the resolution at the expense of sensitivity. Thus, with
$\approx$ 150--200 resolved galaxies out to 
$z<0.09$, MIGHTEE-HI will provide a large homogeneous 
sample beyond the local Universe, which will significantly contribute 
to the above-mentioned studies. Additional information can be gained
from the shape of the resolved \HI\ spectral profile \citep{Watts2020}.

The other caveat for kinematic modelling is the initial estimates 
of the geometry of the disk, that is, the estimates of the position 
and inclination angles. While current kinematic modelling software 
is able to blindly predict the systemic velocity or the position 
of the centre of a galaxy, as well as locate an object in the 
data cube, the results will be unrealistic if an inaccurate initial estimation 
of the position and/or inclination angle is given. However, the
ancillary photometric data available to MIGHTEE-HI will enable initial
estimates of the disk geometry. 

\begin{figure}
\includegraphics[width=\columnwidth]{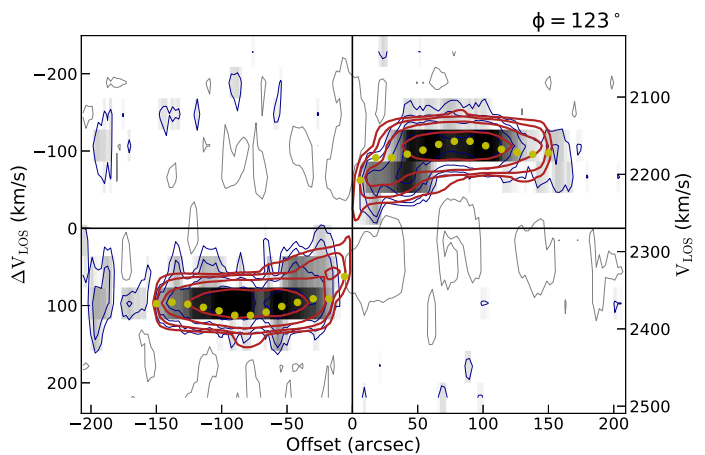}
\caption{Position-velocity diagram of NGC~895, using \HI\ data from
  MIGHTEE-HI, is shown in grey scale and blue 
contours. The 3D kinematic model made with $^{\rm 3D}$Barolo 
\citep{DiTeodoro2015} is shown with red contours, and the resulting 
rotation curve of the galaxy, projected on the data, is shown with 
yellow dots.}
\label{fig:pv895}
\end{figure}

As an example of a well-resolved galaxy, Fig.~\ref{fig:pv895} shows
the position-velocity 
diagram of NGC~895 using data from MIGHTEE-HI, 
 and its kinematic model produced using the $^{\rm 3D}$Barolo software
 \citep{DiTeodoro2015}. The synthesised beam for these observations is
 $11\times 9$ arcsec, while the galaxy is 280 arcsec across the major
 axis. Within the modelling, the separation between the points on the
 rotation curve is 10 arcsec, therefore we model one data point of
 rotation curve per synthesised beam, and the galaxy is $\sim$28 beams
 across the major axis. The resulting rotation velocity of this galaxy is 
$\rm V_{rot}= \pm 165$ kms$^{-1}$, which is consistent with the previous 
measurements of the same galaxy from the VLA observations \citep{Pisano2002}.

\begin{figure}
\includegraphics[width=\columnwidth]{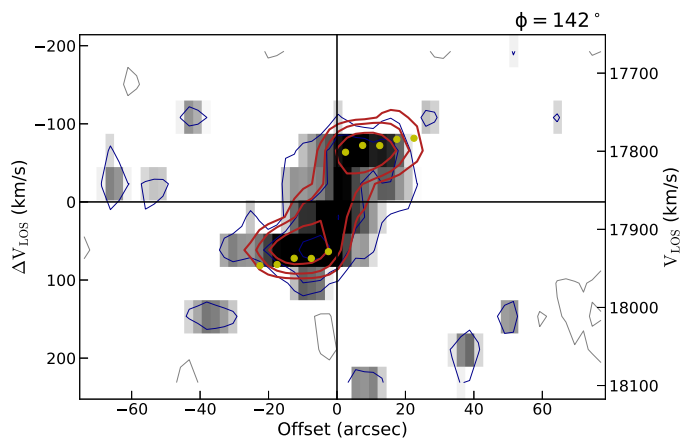}
\caption{Position-velocity diagram of a smaller galaxy detected within
  MIGHTEE-HI, showing that kinematic modelling can be performed
  successfully when the galaxy is only marginally resolved, with three
  resolution elements across the major axis. The MIGHTEE-HI data are
  shown in grey scale and blue contours, while the 3D kinematic model
  is in red contours, and the resulting rotation curve is
  shown with yellow dots.}
\label{fig:smol}
\end{figure}

Initial tests suggest that smaller galaxies can also be
successfully modelled with $^{\rm 3D}$Barolo. The galaxy shown in
Fig.~\ref{fig:smol} is at $z=0.061$, 
and with an \HI\ diameter of 45 arcsec, only spans 3 resolution
elements across its major axis. The synthesised beam here is
$14.5 \times 11$ arcsec. Within $^{\rm 3D}$Barolo, two points of
rotation curve per beam were requested, resulting in a rotation curve
with points separated by 5 arcsec. Initial estimates of the galaxy inclination
and position angles were derived from the optical photometry, but kept
as free parameters within the modelling. The resulting rotational velocity of
the galaxy at the outer most point of the rotation curve is $V_{\rm out} =
123$\,\kms, in good agreement with the rotational velocity
measured at the full-width half maximum (FWHM) of the integrated line
profile ($V_{rot} = 127$\,\kms), corrected for inclination and random motions.

\subsection{The Tully-Fisher relation}\label{subsec:tfr}

The dynamical scaling relations of disk galaxies exhibit 
tight correlations between the main galaxy properties: size, 
mass, and rotational velocity. Among these is the Tully-Fisher 
relation (TFr, \citealt{Tully1977}), and the baryonic TFr, linking the
baryonic content of a galaxy to its 
rotational velocity \citep{McGaugh2000, Ponomareva2018, Lelli2019}. 
Moreover, the TFr holds over a wide wavelength range 
\citep{Ponomareva2017} and in various environments 
\citep{Willick1999}. It has also been studied extensively in the 
theoretical framework of  galaxy formation and evolution 
\citep{Vogelsberger2014,Schaye2015}, as the success 
of any particular theory depends on its ability to reproduce 
the statistical properties of the TFr such as slope, scatter 
and the zero point. 

MIGHTEE-HI will provide a unique opportunity to study 
the statistical properties of the TFr homogeneously at 
various wavelengths. The spectroscopic ancillary data 
will be of great importance to reduce the scatter associated 
with the distance uncertainty. Moreover, MIGHTEE-HI is
a volume-limited, blind survey encompassing various galactic 
environments. This will reduce any selection 
bias that might be present in targeted studies. 

 Comparing the TFr at different redshifts is a powerful tool to
 constrain galaxy growth. Significant evolution in the TFr 
would suggest an imbalance in the accretion histories
 and mass assembly of dark matter and baryons. At low redshifts, 
\HI, resolved rotation curves and/or the velocity width of the \HI\ line are used as the
kinematic tracer, as they provide the best measure of the total 
gravitational potential of a galaxy by extending far beyond
the optical radius into the dark matter halo. However, to date \HI\ has
remained undetectable in emission beyond the local Universe. 
Carbon monoxide (CO) has been used as the kinematic tracer 
at higher redshifts \citep{Topal2018}, but its distribution is known
to be compact and may not probe the dark matter halo potential. The 
same applies to H$\alpha$ and [\ion{O}{iii}]\ lines, which are 
also used as kinematic tracers for the TFr studies at higher 
redshifts (e.g. \citealt{Ubler2017}; \citealt{Tiley2019}). Moreover, current studies are not
conclusive as there is no strong agreement about the presence
or absence of the evolution of the TFr. Thus, MIGHTEE-HI, in
combination with LADUMA, will allow us to study the \HI\ TFr in
redshift bins up to z$\sim$0.4 using direct detections and statistical
methods (see Sect.
 \ref{subsec:stacking}), and compare it to the relation using galaxies
 selected, and with velocities measured, in a consistent way.

\subsection{\HI\ in low-mass galaxies}\label{subsec:lowmass}

Dwarf galaxies can provide critical tests of our cosmological model, and are 
excellent laboratories for studying the processes that control galaxy
formation. Indeed, these two areas of study are inherently connected
as many of the noted discrepancies between observations of dwarf galaxies and
predictions from $\Lambda$CDM simulations are resolved when the galaxy
formation processes, notably the important role of star formation
feedback, are accounted for (see review by \citealt{Bullock2017}).

\HI\ observations offer a powerful tool for identifying isolated,
gas-rich dwarf galaxies and for understanding those dwarf galaxies,
including their internal kinematics and dark matter halo hosts. For
example, several of the lowest-metallicity systems known in the Local
Universe have been discovered via their \HI\ emission
(e.g. \citealt{Skillman2013}; \citealt{Hirschauer2016}), providing a
local opportunity, where high spatial 
resolution can be achieved, to study star formation conditions similar
to those in the early Universe. \HI\ studies of extended, extremely low
surface brightness dwarf galaxies (the so-termed ultra-diffuse
galaxy population), indicate that they do not follow the canonical
baryonic TFr and may have an unusual dark matter content
(\citealt{Leisman2017}; \citealt{Mancera2019}), testing our
understanding of galaxy formation.
Compilations of current \HI\ observations and comparisons to $\Lambda$CDM 
simulations indicate that the kinematics of field dwarf galaxies are
inconsistent with predictions for their expected hosting dark matter
halos (\citealt{Papastergis2015}; \citealt{Papastergis2016}). 

Dwarf galaxies have narrower velocity widths than more massive
galaxies due to lower halo masses. Since \HI\ detection limits depend on
the width of the line, different mass galaxies have different
detection limits. In order to predict the number of dwarf galaxies
expected in the MIGHTEE footprint, we assume a typical velocity width
of 30, 50 and 100 \kms\ for \MHI $<10^7$, $10^7-10^8$, and $10^8-10^9$\,
\Msun. In the
MIGHTEE footprint, we expect to detect $\sim$50 galaxies with \MHI
$<10^8$\,\Msun, and $\sim$250 galaxies in the range $10^8-10^9$\,
\Msun. The additional background Fornax volume starts around the
detection limit for $10^8$\,\Msun, and adds $\sim$100 additional
galaxies in the mass range $10^8-10^9$\,\Msun. These predictions do
not account for cosmic variance; given that dwarf galaxies can only be
detected in limited, nearby volumes, these predictions are approximations
only. 

The resolved detections will display the true power of the MIGHTEE-HI 
data. As discussed in Sect.~\ref{subsec:resolved}, at least three resolution
elements are required to enable kinematic modelling. 
With a 12 arcsec beam, only a few galaxies below $10^9$\,
\Msun\ will be resolved. Re-imaging specific sources at higher angular
resolution after detection at lower angular resolution will provide
more resolved dwarf galaxies. For an 8 arcsec beam, there will be
$\sim$50 resolved galaxies below $10^9$\,\Msun. This is comparable in
size to existing samples of dwarf galaxies in this mass range but
would provide a blind, rather than targeted, sample. The derived
rotation velocities will allow an optically unbiased view of the
baryonic TFr at dwarf galaxy masses plus 
an examination of the `too-big-to-fail' problem in the field
(\citealt{Papastergis2015}; \citealt{Papastergis2016}). 

The multi-wavelength ancillary data will allow galaxy formation processes to be
studied in detail. The \HI\ data provides the gas reservoir, plus a distance, enabling
stellar masses and star formation rates to be derived.
While MIGHTEE-HI will not provide the large numbers of dwarf galaxies
that surveys such as WALLABY and the wide-area tier of the Apertif
project will, the exceptional multi-wavelength data coverage and 
potential for higher spatial resolution will enable more detailed
studies, complementing the work to be done with the wide-area \HI\
surveys. 

\subsection{\HI\ as a function of environment}

The \HI\ content of galaxies is a function of environment on a variety
of scales. On the largest scales, \citet{Jones2018a} showed the effect
of large-scale structure on the determination of the HIMF, with the
largest effect seen at low \HI\ masses. Splitting the full ALFALFA survey
area into a `Spring' region, containing the Virgo cluster and the
Local Supercluster, and a `Fall' region, containing a large void in
front of the Perseus-Pisces supercluster, the faint end slope of the
HIMF differs by more than 3-$\sigma$. This illustrates the need to
encompass a large cosmic volume.

Focusing directly on the Virgo cluster itself, the VLA \HI\ Imaging
survey of Virgo galaxies in Atomic gas (VIVA; \citealt{Chung2009}) 
looked at detailed \HI\ properties of galaxies at a range of
cluster-centric distances. Galaxies near the cluster centre were found
to have small, truncated \HI\ disks with respect to the sizes of their
stellar disks, while galaxies at larger distances had asymmetric \HI\
distributions, pointing away from the cluster centre. The MFS will
extend this work, obtaining similar data for the Fornax cluster in the
Southern Hemisphere. While the MIGHTEE-HI survey volume does not
contain clusters as massive as Virgo, it does contain 10--15
lower-mass clusters and galaxy groups per square degree (see, for
example, the X-ray selected catalogue of \citealt{Finoguenov2007} within the
COSMOS region, and \citealt{Finoguenov2010} and \citealt{Adami2018}
covering the XMMLSS), providing a variety
of environments to explore. At $z<0.1$, as seen in
Fig.~\ref{fig:resolved}, the resolution is sufficient to investigate
the \HI\ morphologies of the \HI -massive galaxies. At higher
redshifts, the integrated \HI\ properties of galaxies, along with
the spectrally resolved profiles, can be investigated in these
different environments.

The group environment, while not as extreme as massive clusters, has
also been shown to affect the \HI\ content of galaxies, with
group halo mass and location within the group (\citealt{Hess2013};
\citealt{Yoon2015}; \citealt{Odekon2016}; \citealt{Brown2017})
influencing the amount of \HI\ detected. This manifests as a
flattening of the faint end slope of the HIMF (\citealt{Pisano2011};
\citealt{Said2019}; \citealt{Jones2020}). It has also been found that cosmic structures
such as filaments can affect the \HI\ content of galaxies
\citep{Odekon2018}. The importance of large-area observations reaching beyond
the nominal group extent is well illustrated in \citet{Serra2013}, who
find a tail of \HI\ clouds containing \MHI\ $=5\times 10^8$\,\Msun\
extending 40 arcmin, postulated to be the result of interactions
between a gas-rich galaxy with the group potential. 

Previous work studying \HI\ in these environments
has all been limited to $z\sim 0$ by the available data. With
the deep imaging available over the full MIGHTEE footprint, along
with highly complete spectroscopy, we will be able to identify the
different galaxy environments, including clusters, groups and filaments,
over the full redshift range of MIGHTEE-HI, and investigate the
behaviour of \HI\ as a function of environment over the past 5
billion years. 

\subsection{The interface between galaxies and the surrounding
  medium}\label{subsec:accretion}

In recent years, it has become clear that galaxies, and in particular
star forming galaxies, cannot be isolated systems as far as their gas
content is concerned. One main indication is that the timescale over
which the star formation in a galaxy would consume all the gas in that
galaxy is much shorter than a Hubble time
(e.g. \citealt{Bigiel2008}). If there is no channel though which
star forming galaxies acquire new gas, the present gas reservoirs
would be depleted in only a few Gyr. Studies of our own Galaxy show
that the disk stars have formed at a more or less constant rate for
the last 12 Gyr (e.g. \citealt{Schoenrich2009}), which can only occur
when the gas content of the Milky Way is also more or less
constant. Chemical evolution models of the stellar abundances also
indicate extensive and continuous accretion of fresh unpolluted gas
(e.g. \citealt{Chiappini2001}). 

It is still unclear how star forming galaxies acquire fresh gas in
order to maintain their star formation rate. However, there are several
possibilities for how galaxies could be doing this. One obvious
candidate is the accretion of gas-rich companions, while another
option is that the accretion of gas structures similar to the Galactic
High-Velocity Clouds plays an important role. This latter process could
correspond to the so-called cold accretion, which is seen in modern
numerical simulations of galaxy formation. Both types of accretion
are observed in some nearby spiral galaxies, but the data suggest that
the observed accretion rates may not be sufficient to maintain star
formation at current levels, unless the observed cold gas clouds that
are accreting have significant ionised counterparts \citep
{Sancisi2008,DiTeodoro2014,Putman2012}. 

An alternative to accretion of external gas is that gas from the hot
gaseous halo, which spiral galaxies are known to possess, cools and
rains down onto the disk. This can help rebuild a substantial
  disk after an early merger of two disk galaxies 
\citep{Athanassoula2016}. Additionally, an interesting idea that has
been proposed is that galactic fountains could play an important role
in stimulating cooling of the hot halo gas 
\citep{Fraternali2008}. The energy released by supernovae and
stellar winds drives relatively cold gas flows into the gaseous halo
where they interact with the hot, low-metallicity halo gas. The mixing
of these two gas components drastically reduces the cooling time of
the hot halo gas so that part of the halo gas cools and rains down
onto the disk, adding to the existing gas reservoir. Observations have
given indications that such an interface layer exists in at least some
galaxies. Detailed \HI\ observations of a small number of nearby
galaxies (e.g.\ NGC 2403 \citealt{Fraternali2001}; NGC 891
\citealt{Oosterloo2007}) and more recently the result from the HALOGAS
survey \citep{Heald2011}, show that the thin \HI\ disk of spiral
galaxies is surrounded by a thicker \HI\ disk, which possibly corresponds
to the layer where the interaction between the galactic fountain and
the halo occurs. These thick \HI\ disks have different kinematics
compared to the thin disks: They have lower rotation speeds and also
display a small, inward motion \citep{Marasco2019}. These
characteristics are consistent with the galactic fountain models. 

However, sufficiently sensitive \HI\ data that allow us to study gas
accretion, and galactic fountains in detail, are available only for a
small, not entirely representative sample of galaxies and there are
still several important open questions. The HALOGAS sample from
\citet{Marasco2019} only contains 15 galaxies and is the largest
collection of homogeneously studied objects. 
The MIGHTEE-HI survey has the required sensitivity and resolution to
increase the number of galaxies for which data are available. In
particular, the galaxies will not be pre-selected, allowing a blind
census of gas accretion and of the role of galactic fountains play.

\subsection{\HI\ below the detection limit}\label{subsec:stacking}

Science from MIGHTEE-HI is not restricted to direct detections of \HI\
in galaxies. Statistical techniques can be employed to extract
information from below the noise floor, to extend to lower \HI\ masses
and higher redshifts. The most common method for
determining the average \HI\ properties of an ensemble of galaxies is
spectral stacking, where one relies on the existence of sufficiently
accurate optical spectroscopy of the galaxies under investigation in
order to align the \HI\ spectra of 
a large number of non-detections, $N$, into a common reference frame, and
then assume the noise of the co-added spectrum decreases as
$N^{-1/2}$.

This has been successfully attempted over large areas
\citep{Delhaize2013}, as a function of galaxy type \citep{Brown2015},
and as a function of stellar mass \citep{Healy2019}. \citet{Meyer2016}
also showed the power of spectral stacking to recover the statistical 
properties of the TFr. Regions of the
sky with either large numbers of galaxies, such as clusters (although
these regions are biased by design)
\citep{Jaffe2016}, or regions with particularly dense optical spectroscopic
coverage \citep{Rhee2016}, are also good candidates for spectral
stacking studies. The expanded frequency range of next-generation
facilities enables \HI\ detections to increasing redshifts
\citep{Bera2019}. A new technique employing Bayesian statistics to
model the distribution of fluxes from within a data cube has been shown to
directly recover the HIMF from simulated data, rather than infer it
from average values, extending below the formal noise
limit \citep{Pan2020}. As described in Sect.~\ref{subsec:ancdata},
the MIGHTEE fields already have extensive optical spectroscopy, with
more to come, which has been shown to be essential for the success of
any stacking experiment \citep{Maddox2013}.

\subsection{\HI\ intensity mapping}

Neutral hydrogen intensity mapping (\HI\ IM) has been proposed as a
new technique to probe the large-scale structure of the Universe and
deliver precision constraints on cosmology \citep{Bharadwaj2001,
  Battye2004, Loeb2008}. The idea is to
integrate all of the \HI\ emission within a given pixel, of specified
angular and frequency resolution. The advantage is that one can
achieve higher sensitivity as we do not require a detection of
individual galaxies (i.e. there is no threshold for detection). One of
the main issues is to disentangle the desired signal from the smooth
but strong foreground contaminants and instrumental effects
\citep{Wolz2014, Olivari2018}.

Several experiments have been proposed in order to measure this
signal, using single dish telescopes or interferometers
\citep{Bandura2014, Xu2015,
  Newburg2016, Santos2015}. A precursor
survey to the SKA with MeerKAT has also been proposed using the
single dish measurements \citep{Santos2017}.

Measurements using the Green Bank Telescope produced the first tentative
detection of the cosmological \HI\ intensity signal by cross-correlating
with the WiggleZ redshift survey \citep{Chang2010,Switzer2013}.
More recently, a survey using the Parkes telescope made a detection in
cross-correlation with the 2dF survey \citep{Anderson2018} at $z <
0.098$. They measured an amplitude of the cross-power spectrum lower
than expected from a dark matter power spectrum, assuming an \HI\ bias
and mass density equal to measurements from the ALFALFA survey
(especially at $k \sim 1.5 {\rm h/Mpc}$). This indicates a possible
lack of clustering or a small correlation coefficient with the optical
galaxies that needs to be further investigated. 

In addition to measuring the auto correlation, using the data from
MIGHTEE, we plan to search for the \HI\ 
IM signal in cross-correlation with optical galaxy catalogues, taking
advantage of the excellent spectroscopic and photometric coverage of
the survey fields. We note that such a detection would be a first for an
interferometer. 

Within a single MIGHTEE pointing, we will be able to probe scales
between 0.3 ${\rm Mpc}^{-1} < k < $ 20 $ {\rm Mpc}^{-1}$ (0.3 to 21
comoving {\rm Mpc}) at $z\sim 0.27$, and using data from multiple MIGHTEE
fields reduces the cosmic variance of the power spectrum. Although
these scales are in the quasi-linear regime (i.e. smaller scales than
the baryon acoustic oscillation window), they can still provide
important cosmological constraints. Such measurements would provide a
wealth of information for comparison with \HI\ numerical simulations
\citep[e.g.][]{Dave2019}. In addition, these measurements
will help constrain $\Omega_{\rm HI}$ and, indirectly, the HIMF down
to very low mass scales, for which a measurement of 
the shot noise will also help (e.g. \citealt{Wolz2017}), and would not
only include optically selected galaxies, but all the \HI\ emission,
irrespective of the galaxy with which it is associated.  

\subsection{Scaling relations}

As the \HI\ content of galaxies is a dynamic quantity, it is crucial
that we are able to observe it as a function of cosmic time. The existing
large-area surveys, such as HIPASS and ALFALFA, provide excellent
$z=0$ reference points for the current \HI\ content of galaxies. The
Analysis of the interstellar Medium in Isolated GAlaxies project
(AMIGA; \citealt{Lourdes2005}) provides specific scaling relations at
$z=0$ derived from \HI\ spectra as a reference for environmental
effects \citep{Jones2018b}. With MIGHTEE, we have the
opportunity to extend these observations to lookback times of up to 5
Gyr. With \HI\ serving as one source of 
raw fuel for star formation, and thus the build-up of stellar mass,
incorporating the neutral hydrogen component into our multi-wavelength
census of galaxy evolution is essential.

The relation between the \HI\ content of galaxies and their stellar
mass at $z\sim 0$ is known to be non-linear, and a function of the
dark matter halo parameters \citep{Maddox2015}. With MIGHTEE-HI, we
can measure the \HI\ mass of galaxies as both a
function of stellar mass and redshift.
The quality of the ancillary data available in the MIGHTEE footprint is
essential, as it allows us to use statistical methods to probe scaling
relations beyond where \HI\ is directly detected, extending to higher
redshifts and lower masses. 

As MIGHTEE is a commensal spectral line and continuum survey, we have
deep radio continuum images over the same area, from which we can measure
star formation rates (SFRs) unaffected by dust obscuration, as well as
active galactic nucleus (AGN)
activity. Using the conservative 2$\mu$Jy beam$^{-1}$ root mean square
(RMS) for MIGHTEE,
we can convert this to a 5-$\sigma$ limiting SFR as a function of
redshift using the relation from \citet{Murphy2011}. At low redshifts,
$z<0.09$, very low SFRs are probed, down 
to 0.15\,\Msun\ year$^{-1}$. At the high-redshift limit probed by \HI\,
$z\sim 0.58$, we are sensitive to SFR$\sim$10\,\Msun\ year$^{-1}$.

\subsection{\HI\ in galaxy evolution simulations}

The baryon cycling paradigm of galaxy evolution asserts
that gas flows into and out of galaxies govern their evolution and
properties~\citep{Somerville2015}.  \HI\ plays a key role in this
by tracing gas around galaxies that is the reservoir of future star
formation~\citep{Crain2017} as well as cool gas
outflows~\citep{FaucherGiguere2016}.  Numerical simulations of
galaxy formation and evolution have shown that, even with many
different choices for feedback models, it is possible to broadly
reproduce observed stellar, star formation rate, and black hole
properties over a range of redshifts (e.g. EAGLE, IllustrisTNG,
and {\sc Simba}; \citealt{Schaye2015,Pillepich2018,Dave2019}).  However, the
predicted properties of cold gas in and around galaxies vary widely,
particularly in terms of their evolution~\citep{Dave2020}.

MIGHTEE-HI will provide an unprecedented sample of observations for
testing models at the present epoch out to intermediate redshifts.  In
particular, the large area will enable homogeneous environmental
studies to better constrain the processes by which galaxies lose
their \HI\ within dense regions, which is important for
quenching star formation in both central and satellite galaxies.  Feedback and gas
stripping processes, for instance, appear to be crucial for solving
the  too-big-to-fail problem \citep{Verbeke2017,
Oman2019a}, and for predicting \HI\ removal and consumption
timescales in dense
environments~\citep[e.g.][]{Marasco2016,Mika2019}.  Indeed,
the suppression of \HI\ likely extends into shock-heated filamentary
structures well outside groups and clusters~\citep{Wang2015,Mika2018},
and the multi-wavelength
data associated with MIGHTEE will be crucial for understanding the
association of \HI\ and star formation in the earliest stages of
environmental quenching.  Finally, the dynamical information provided
by the high velocity resolution of MeerKAT will enable careful
studies of the connection between galaxies and halos via the baryonic
TFr (Sect.~\ref{subsec:tfr} and e.g. \citealt{Glowacki2020}).

It is becoming routine to generate \HI\ data cubes from simulations
including instrumental effects (e.g. MARTINI\footnote{\url{
https://kyleaoman.github.io/martini/build/html/includeme.html}};
\citealt{Oman2019b}),
which will enable closer and more robust comparisons between both
cosmological and zoom simulations, versus MIGHTEE-HI and ancillary
data.  Thus there is great potential for synergistic studies
between cosmological hydrodynamical simulations and the \HI\ studies
of galaxies in the MIGHTEE-HI project, in order to constrain some
of the most uncertain physical processes of galaxy formation associated
with the baryon cycle.


\section{Survey progress}\label{sec:progress}

Early Science observations with MeerKAT were undertaken from
mid-2018 through 
mid-2019. For MIGHTEE, Early Science employs the full 64-dish array, but 
with limited spectral resolution, 208\,kHz (44\,\kms\ at $z=0$) channels instead of 
26\,kHz (5.5\,\kms) resolution. MIGHTEE has Early Science data for the COSMOS
field, and three overlapping pointings within the XMMLSS field. There
are three tracks (16 hours on source) for COSMOS, and two 8-hour
tracks for each of the XMMLSS pointings (12 hours on source). Full
science observations, with 26kHz spectral resolution, have commenced
in 2020. The Early Science data have been reduced and spectral cubes
generated. See Frank et al. (in prep.) for a full description of the
MIGHTEE-HI Early Science Data.

\subsection{Early Science}\label{subsec:highlights}

Within the restricted area observed in Early Science, there are
already enough detections spanning a range of redshift and \MHI\ for
population studies to begin, along with individual objects that
warrant detailed attention. We highlight one of them
here, to demonstrate the advantages of the MeerKAT instrument, and the
discovery potential of the full survey. 

\begin{figure}
\includegraphics[width=\columnwidth]{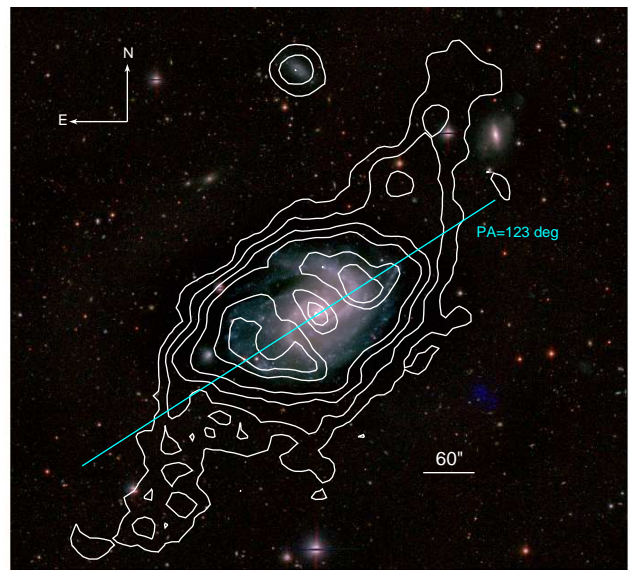}
\caption{MIGHTEE-HI \HI\ total intensity contours, overplotted on the
  HSC $grY$ image of NGC~895. Contours are at $1\times 10^{20}$,
  $3\times 10^{20}$, $6\times 10^{20}$, $1\times 10^{21}$, $2\times
  10^{21}$, and $2.5\times 10^{21}$ atoms cm$^{-2}$. The contours in
  the galaxy centre indicate a depression of \HI, also seen in
  Fig.~\ref{fig:pv895}. The \HI\ contours have been smoothed to
  $20\times 20$ arcsec, and are based on $\sim$\,12 hours of on-source
  integration. The scale-bar shows 60 arcsec, while the synthesised
  beam is $11\times 9$ arcsec. The position angle from the modelling
  in Fig.~\ref{fig:pv895} is marked in cyan.}
\label{fig:mom0895}
\end{figure}

As described in Sect.~\ref{subsec:resolved}, several galaxies within
MIGHTEE-HI will be well-resolved by MeerKAT. Figure~\ref{fig:pv895}
illustrates the kinematic modelling possible for such objects, while
Fig.~\ref{fig:smol} shows that kinematic modelling can also be
performed on galaxies that are only marginally resolved. The full
investigation of the kinematics of MIGHTEE-HI Early Science galaxies
is underway and will appear in future works. Figure~\ref{fig:mom0895}
shows the total intensity, or moment 0, contours of NGC~895
overplotted on the HSC $grY$ image. This one system illustrates
several advantages of the high sensitivity, high-resolution MeerKAT
data. Visible in Fig.~\ref{fig:mom0895} are not only the \HI\
associated with the galaxy disk at $z=0.0075$, but also tidal
extensions, along with a dwarf companion also clearly detected. The
mass of the main galaxy is \MHI\,$=1.7\times10^{10}$\,\Msun, similar
to the \MHI\,$=1.3\times10^{10}$\,\Msun\ found in
\citet{Pisano2002}. The deep HSC imaging clearly shows the dwarf
companion stellar counterpart, which had no catalogued redshift until
the \HI\ observations confirmed its association with the main
galaxy. The companion mass is \MHI\,$=5\times10^7$\,\Msun.

\subsubsection{Catalogue construction}\label{subsec:catalogue}

With such a rich data set, and the large number of \HI\ detections, the
data will be collated within the MIGHTEE-HI working group into
catalogues. In the first instance, source finding is performed
visually and unguided, meaning no prior information about the location
of galaxies is used. This will be supplemented by including automated source
finding algorithms. We employ The Cube Analysis and Rendering Tool
for Astronomy (\texttt{CARTA}\footnote{\url{https://cartavis.github.io/}})
for cube inspection. Cubes at multiple spatial and spectral resolutions can be created,
according to specific science cases. For example, low-mass galaxies
will require high spectral resolution to resolve the narrow line
widths, whereas column density sensitivity for low surface brightness features
can be enhanced at the expense of spatial resolution. 

Once the position and frequency of each \HI\ detection is known, cubelets
around the sources are extracted from the cube. These cubelets are
processed by the source finding and parameterisation tool \texttt{SOFIA}
\citep{Serra2015}, where masks around emission in the relevant
spectral channels are created. From these masks, the \HI\ flux is
summed, and the central frequency determined. With the relatively
coarse spectral resolution of Early Science data, the central
frequencies of galaxies are limited in precision, but will improve
with full spectral resolution data. The \HI\ flux combined with the
emission frequency can be converted into an \HI\ mass. 
Other data products of interest are the integrated \HI\ line profiles,
the width of the \HI\ emission line at half the peak flux, $W_{50}$,
the integrated flux map (moment 0) and velocity map (moment 1). These
data products will all be made available first to the MIGHTEE-HI
working group, and then publicly released after the data proprietary
period has expired, via staged data releases coordinated across the
MIGHTEE working groups. The Early Science Data Release will be
mid-2021, with reduced data products hosted on the
IDIA\footnote{\url{https://www.idia.ac.za/}} facility. 

From the extensive ancillary data within the MIGHTEE survey footprint
described in Sect.~\ref{subsec:ancdata}, several value-added
quantities are available. The spectroscopic redshifts for all the \HI\
sources are by definition given from the \HI\ data themselves, as
MIGHTEE is in fact a spectroscopic survey. This is particularly
useful for the low-mass dwarf galaxies and low surface brightness
galaxies, for which acquiring an optical redshift is difficult and
thus these types of galaxies are often not targeted in optical surveys.
If an optical spectrum also exists, the SFR can
be derived directly from suitable emission lines, such as
  H$\alpha$. Additionally, a 
dust-free estimate of the SFR can be derived from the MIGHTEE radio
continuum map.

Using the available deep multi-wavelength optical and NIR imaging, we
perform resolved photometry on the \HI -detected
galaxies. An elliptial aperture is determined for each galaxy individually,
encompassing as much of the galaxy light as possible while excluding
the light of foreground, background or neighbouring objects. The
aperture is defined on the HSC $G$-band image and transferred directly
to the CFHT $u$-band, HSC $riZY$ bands, and NIR $JHKs$ bands,
and the aperture photometry extracted, resulting in 9-band photometry. 
Coupled with the accurate redshifts provided by the \HI\
line, the photometry is fit by the spectral energy distribution (SED)
fitting code \textit{Le PHARE} (\citealt{Arnouts1999}; \citealt{Ilbert2006}) to
produce stellar masses, star formation rates, and stellar ages. Thus,
for each \HI -selected source, we have coordinates, a spectroscopic
redshift, and the \HI\ properties of the galaxy, along with the optical properties of
the stellar component, including stellar mass and star formation
rate. 

\subsection{MIGHTEE-HI galaxies}\label{subsec:family}

\begin{figure}
\includegraphics[width=\columnwidth]{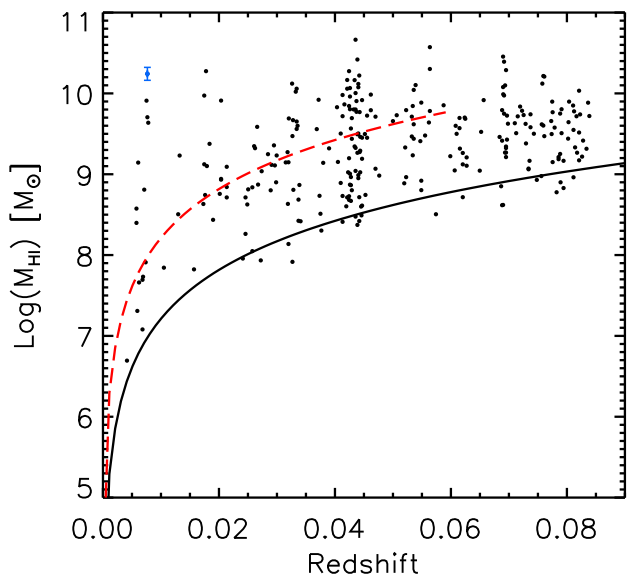}
\caption{\HI\ masses as a function of redshift for the 
  Early Science \HI\ detections. Black points mark individual
  objects, while the curved black line indicates the 5-$\sigma$ flux
  limit used for the simulations. An estimate of the uncertainty on the \HI\
  masses is illustrated for the blue point with error bars in the top left
  corner. The red dashed line is the 
  approximate flux limit of the ALFALFA survey, which does not extend
  beyond $z\sim 0.06$. MIGHTEE-HI Early Science detections are found all the way to
  the upper redshift limit of $z=0.084$.}
\label{fig:MHI_z}
\end{figure}

While several individually detected galaxies warrant close
investigation, the sample of \HI -detected galaxies is already large
enough to enable ensemble studies. 
Figure~\ref{fig:MHI_z} shows the range of \HI\ masses of galaxies
visually detected in cubes as a function of redshift. 
The solid black line in the figure is the 5-$\sigma$ flux 
limit from the simulations described in Sect.~\ref{subsec:sims},
altered to better represent the Early Science integration time and
channel width, indicating the expected mass sensitivity limit as a function of
redshift. The MIGHTEE-HI flux limit is a factor of ten deeper than the
approximate flux limit of the large-area ALFALFA survey,
shown as a red dashed curve, and extends further in redshift.
Data at frequencies below 1310\,MHz, corresponding to
$z>0.084$, have not yet been inspected and are not shown here.
Isolated galaxies (i.e. not obvious satellites of larger
galaxies) with \MHI\ as low as $10^7$\,\Msun\ have 
been detected, in line with predictions. We expect the number of very low mass
detections to increase with improved spectral resolution, as the
intrinsic linewidths of these low-mass galaxies can be narrower than
the Early Science 44\,\kms\ channels.

From the noise properties of the cubes, we can derive a
measure of the statistical uncertainty on the \HI\ masses of our
individual objects. For each galaxy, the local RMS in emission-free
regions is measured and converted to an \HI\ mass error, assuming no
uncertainty on the distance. At large masses, the uncertainty is
$\sim$5 per cent, and at lower masses, $\sim 10^8$\,\Msun, increases to
approximately 20 per cent. The uncertainty typical for a low-mass
galaxy is illustrated by the error bars on one point in the top left
corner of Fig.~\ref{fig:MHI_z}. 

Environmental studies benefit from the large FoV of MeerKAT. Galaxy
groups uncovered in the Early Science data, as traced by their \HI\
detections, span tens of arcminutes and fit easily into a single
MeerKAT pointing. An overdensity is already apparent in
Fig.~\ref{fig:MHI_z} at $0.04<z<0.045$ within the XMMLSS field.
At low redshift, $z\sim 0.04$, 60 arcmin corresponds
to a physical scale of nearly 3 Mpc. At $z\sim 0.4$, the upper redshift
limit of where we expect to directly detect \HI, 60 arcmin corresponds
to 19 Mpc, large enough to trace large-scale structure.

\subsubsection{Scaling relations}

\begin{figure}
\includegraphics[width=\columnwidth]{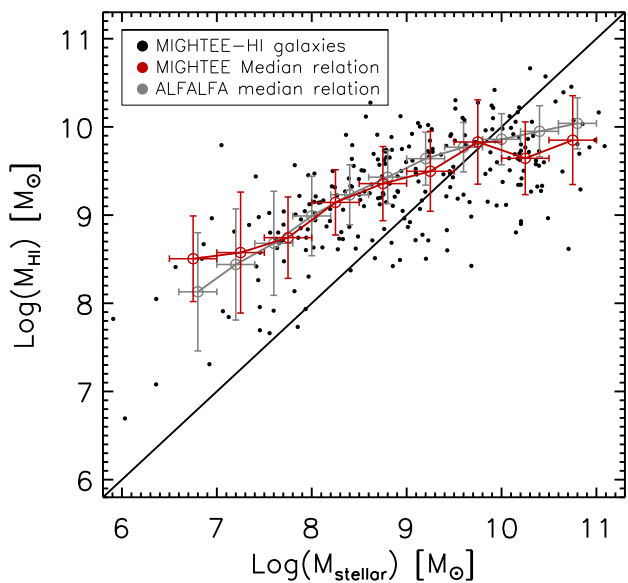}
\caption{\HI\ masses as a function of stellar mass for MIGHTEE Early
  Science \HI\ detections, spanning $0<z<0.84$. Black points
  mark individual objects. The curved red line indicates the median and
  1-$\sigma$ uncertainty of the
  relation from the MIGHTEE-HI points, while the grey line is the same
  relation derived from the ALFALFA survey \citep{Maddox2015}. The
  horizontal error bars indicate the width of the stellar mass bins.}
\label{fig:MHI_Mstellar}
\end{figure}

The \HI\ mass of the Early Science galaxies as a
function of stellar mass is shown in Fig.~\ref{fig:MHI_Mstellar}. The
\HI -detected galaxies span 5 orders of magnitude in stellar mass,
reaching $<10^7$\,\Msun. The deeper optical imaging available within the
MIGHTEE footprint allows us to probe lower stellar masses than
ALFALFA, which relied on the shallower SDSS imaging. The
  low-mass end, below \Mstellar $<10^7$\,\Msun, is not yet well
  populated, but will become increasingly so as more area is surveyed.

The grey line in Fig.~\ref{fig:MHI_Mstellar} traces the median
relation derived from the 40\% ALFALFA catalogue in bins of width
Log(\Mstellar)=0.4, along with illustrative 1-$\sigma$\ uncertainties,
as determined in \citet{Maddox2015}. The red line traces the median
relation derived from the 5 deg$^2$ of MIGHTEE-HI Early Science data,
in bins of Log(\Mstellar)=0.5, also with illustrative 1-$\sigma$\
uncertainties. The deficit seen by the MIGHTEE-HI data at the
high-mass end may be a result of the limited volume probed in Early
Science. However, at intermediate masses, the MIGHTEE-HI and
ALFALFA relations are in very good agreement.
The apparent break at \MHI\,$\sim$10$^9$\,\Msun\ in this
relation was first noted in the ALFALFA data \citep{Maddox2015}, and
was not seen in HIPASS data, nor any galaxy simulations
available. MIGHTEE-HI provides the first independent confirmation of
the shape of the relation, employing data that extend a factor of ten
deeper in \HI\ mass. Studies of scaling relations illustrate the need
for a combination of sensitivity, required to detect the low-mass systems, large area
coverage of the MIGHTEE survey, to encompass a larger search
volume, and the correspondingly deep optical and NIR imaging, to investigate
the faint stellar counterparts.

\section{Summary}\label{sec:summary}

The next generation of radio facilities are rapidly becoming
operational, probing parameter space that has until recently remained largely
unexplored. With the exceptional sensitivity of the SKA precursor
MeerKAT, large-area surveys are able to reach unprecedented depths and
redshifts. The MeerKAT MIGHTEE LSP is well underway, with several
square degrees of Early Science data in hand. The survey design and scientific
motivation for the \HI\ component of the MIGHTEE LSP, designated as
MIGHTEE-HI, are outlined here. A description of the data products that will
be made available to the LSP, and ultimately, the global
community is given, along with first science results highlighting the
strengths of the survey. The combination
of the new MeerKAT \HI\ data, with the existing 
extensive multi-wavelength information available, provides a powerful
resource for undertaking the science described in
Sect.~\ref{sec:science} and beyond. These new facilities are
set to transform our view of the neutral gas component of galaxies,
adding this crucial ingredient into our descriptions of galaxy
formation and evolution. MeerKAT in particular, with its
  unprecedented combination of 
  sensitivity, resolution, FoV, and bandwidth, is uniquely
  positioned to make great strides in \HI\ science, well in advance of
construction of the SKA.


\begin{acknowledgements}
We thank the referee, Prof Albert Bosma, for his careful reading of
the manuscript and helpful comments which improved this paper.
The authors gratefully acknowledge input from the MIGHTEE-HI working
group for useful discussions and comments on early drafts of this paper.
The MeerKAT telescope is operated by the South African Radio Astronomy
Observatory, which is a facility of the National Research Foundation,
an agency of the Department of Science and Innovation. We acknowledge
use of the Inter-University Institute for Data Intensive Astronomy
(IDIA) data intensive research cloud for data processing. IDIA is a
South African university partnership involving the University of Cape
Town, the University of Pretoria and the University of the Western
Cape. The authors acknowledge the Centre for High Performance
Computing (CHPC), South Africa, for providing computational resources
to this research project.

NM acknowledges support from the Bundesministerium f{\"u}r Bildung und
Forschung (BMBF) D-MeerKAT award 05A2017. BSF, MJJ and IH would like
to acknowledge support from the Africa-Oxford Visiting Fellows
Programme. MJJ, AAP, RAAB and IH acknowledge support from the Oxford Hintze
Centre for Astrophysical Surveys which is funded through generous
support from the Hintze Family Charitable Foundation and the award of
STFC consolidated grants (ST/S000488/1 and ST/N000919/1). NJA
acknowledges funding from the Science and Technology Facilities
Council (STFC) Grant Code ST/R505006/1. RAAB is supported by the
Glasstone Foundation. EAKA is supported by the
WISE research programme, which is financed by the Netherlands
Organisation for Scientific Research (NWO). 
RD acknowledges support from the Wolfson Research Merit Award program
of the U.K. Royal Society. MGS acknowledges support from the South
African Radio Astronomy Observatory (SARAO) and the National Research
Foundation (Grant No.~84156). IP acknowledges support from the Italian
Ministry of Foreign Affairs 
and International Cooperation, Directorate General for the Country
Promotion (Bilateral Grant Agreement ZA18GR02), and from the South
African Department of Science and Technology’s National Research
Foundation (DST-NRF Grant Number 113121) as part of the ISARP
RADIOSKY2020 Joint Research Scheme. RKK, SHAR and WM are supported
through the South African Research Chairs Initiative of the Department
of Science and Technology and National Research Foundation. 

This research has made use of NASA's Astrophysics Data System
Bibliographic Services. This research made use of
Astropy,\footnote{\url{http://www.astropy.org}} a community-developed
core Python package for Astronomy. 
\end{acknowledgements}

%
%


\end{document}